\documentclass[10pt,conference]{IEEEtran}
\IEEEoverridecommandlockouts

\usepackage{cite}
\usepackage{amsmath,amssymb,amsfonts}
\usepackage{graphicx}
\usepackage{textcomp}
\usepackage{xcolor}
\def\BibTeX{{\rm B\kern-.05em{\sc i\kern-.025em b}\kern-.08em
    T\kern-.1667em\lower.7ex\hbox{E}\kern-.125emX}}

\usepackage[bookmarks=false]{hyperref}
\usepackage{booktabs}
\usepackage{algorithm}
\usepackage{algpseudocode}
\usepackage{color}
\usepackage{colortbl} 
\usepackage{comment}
\usepackage[inline]{enumitem}
\usepackage{epsfig}
\usepackage{etoolbox}
\usepackage{fancybox}
\usepackage{float}
\usepackage{hyperref}
\usepackage{listings}
\usepackage{multirow}
\usepackage{multicol}
\usepackage{placeins}
\usepackage{rotating}
\usepackage{setspace}
\usepackage{threeparttable}
\usepackage{times}
\usepackage{verbatim}
\usepackage{wrapfig}
\usepackage[usenames,dvipsnames]{xcolor}
\usepackage{xspace}
\usepackage{mathrsfs} 
\usepackage{fancyhdr} 
\usepackage{pifont} 
\usepackage{booktabs}
\usepackage{bbding}
\usepackage{numprint} 
\usepackage{orcidlink}

\DeclareRobustCommand*{\IEEEauthorrefmark}[1]{%
    \raisebox{0pt}[0pt][0pt]{\textsuperscript{\footnotesize\ensuremath{#1}}}}

\begin{document}
\title{\textit{RoboECC}: Multi-Factor-Aware Edge-Cloud Collaborative Deployment for VLA Models}

\author{\IEEEauthorblockN{
Zihao Zheng\IEEEauthorrefmark{1}, 
Hangyu Cao\IEEEauthorrefmark{2}, 
Jiayu Chen\IEEEauthorrefmark{1}, 
Sicheng Tian\IEEEauthorrefmark{3}
Chenyue Li\IEEEauthorrefmark{2},
Maoliang Li\IEEEauthorrefmark{1},
Xinhao Sun\IEEEauthorrefmark{4},\\
Guojie Luo\IEEEauthorrefmark{1},
$^{\dagger}$Xiang Chen\IEEEauthorrefmark{1}
} 
\IEEEauthorblockA{\IEEEauthorrefmark{1}School of Computer Science, Peking University, Beijing, China} 
\IEEEauthorblockA{\IEEEauthorrefmark{2}School of Computer Science, South China University of Technology, Guangzhou, China} 
\IEEEauthorblockA{\IEEEauthorrefmark{3}School of Artificial Intelligence, Beijing Normal University, Beijing, China} 
\IEEEauthorblockA{\IEEEauthorrefmark{4}School of EECS, Peking University, Beijing, China}
}



\maketitle

\begin{abstract}
\label{tex:abstract}
Vision-Language-Action (VLA) models are mainstream in embodied intelligence but face high inference costs. Edge-Cloud Collaborative (ECC) deployment offers an effective fix by easing edge-device computing pressure to meet real-time needs.
However, existing ECC frameworks are suboptimal for VLA models due to two challenges: (1) Diverse model structures hinder optimal ECC segmentation point identification; (2) Even if the optimal split point is determined, changes in network bandwidth can cause performance drift.
To address these issues, we propose a novel ECC deployment framework for various VLA models, termed \textit{RoboECC}.
Specifically, we propose a model-hardware co-aware segmentation strategy to help find the optimal segmentation point for various VLA models.
Moreover, we propose a network-aware deployment adjustment approach to adapt to the network fluctuations for maintaining optimal performance.
Experiments demonstrate that \textit{RoboECC} achieves a speedup of up to 3.28x with only 2.55\%$\sim$2.62\% overhead.
\end{abstract}

\section{\textbf{Introduction}}
\label{tex:introduction}

Vision-Language-Action (VLA) models have emerged as the mainstream paradigm for Embodied Artificial Intelligence. 
Via diverse encoders, they fuse multi-modal information, generate concrete actions, and exhibit high task success rates as well as robust generalization in robotic control scenarios~\cite{rt2, openvla}.

\begin{figure*}[!t]
    \centering
    \includegraphics[width=7in]{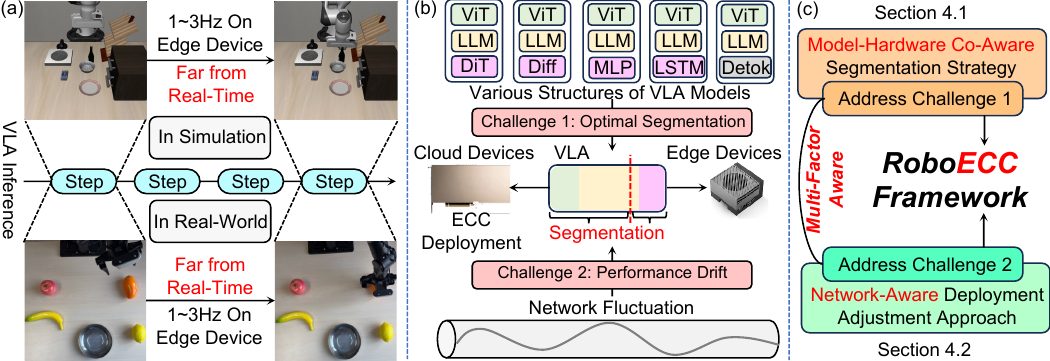}
    \vspace{-2mm}
    \caption{(a) VLA Inference on Edge Devices (b) Challenges of VLA ECC Deployment (c) Overview of the Proposed \textit{RoboECC} Framework}
    \vspace{-4mm}
    \label{fig:1}
\end{figure*}

VLA models incur high inference costs, which severely impede their real-world deployment. 
Unoptimized VLAs only reach an action control frequency of 1$\sim$3 Hz on edge devices (Fig.~\ref{fig:1}~(a)), far below the $\sim$30Hz real-time requirement~\cite{kerv, zheng2026dyq}. 
Even with acceleration methods, this frequency only increases to 10$\sim$15Hz~\cite{heisd} (on consumer-grade GPUs) and drops further on resource-limited edge devices.

Edge-Cloud Collaboration (ECC) is a highly effective deployment strategy for large-scale models~\cite{rapid}. 
Typically, it deploys a model’s backbone on the cloud (with abundant computing resources) and a lightweight segment on edge devices; via high-bandwidth networks, this setup enables collaborative inference~\cite{edgeshard, coedge}. 
This configuration not only eases edge devices’ computing burden but also boosts overall inference speed and service quality.

Effective integration of ECC with VLA models can enable real-time action generation, yet this process is non-trivial.
We summarize two key challenges (Fig.~\ref{fig:1}~(b)).
\textbf{Challenge 1: Due to the component heterogeneity inherent in VLA models, existing ECC frameworks are suboptimal in model segmentation.}
With technological progress, VLA models have developed diverse architectures.
Early models relied solely on detokenizers for token-to-action conversion, while later iterations evolved into specialized Action Models (AM) for action generation—including MLPs, LSTMs, diffusion models, and Diffusion Transformers (DiTs). 
Owing to the distinct computing/bandwidth needs of VLA components and such architectural complexity, existing static ECC frameworks struggle to find the VLA models’ optimal segmentation point.

\textbf{Challenge 2: In end-cloud scenarios, internet bandwidth is subject to constant fluctuations. Even when an optimal segmentation point is set, these bandwidth variations can still lead to suboptimal latency.}
The data transfer volume at the model’s optimal split point reflects only ideal bandwidth conditions. Fluctuations in internet bandwidth can increase or decrease network transmission latency, which in turn alters overall latency. This shifts the optimal split point, requiring fine-grained adjustment.
Existing ECC frameworks can not achieve these, leaving it a challenge.

In this paper, we first develop some case studies to prove these two challenges and conclude design insights.
Based on these, we propose an end-to-end multi-factor-aware ECC framework for optimal VLA deployment, which is named \textit{RoboECC} (Fig.~\ref{fig:1}~(c)). 
Specifically, \textit{RoboECC} consists of two parts.
First, we introduce a model-hardware co-aware segmentation strategy to help find the optimal segmentation point for various VLA models. 
We establish an effective hardware model and VLA structure model, and utilize an automatic algorithm to find the segmentation point.
Second, we introduce a network-aware adjustment method to adapt to the network bandwidth fluctuations and maintain optimal performance.
We build up a parameter-sharing pool for fine-tuning the edge-cloud deployment and finish the corresponding adjustment methods.
Overall, our contributions are threefold:

\begin{itemize}[leftmargin=*]
    \item[$\bullet$] We develop case studies to conclude the key problems of achieving ECC deployment for various VLA models and bring key insights for the ECC framework design.
    \item[$\bullet$] We propose \textit{RoboECC}, a multi-factor-aware ECC deployment framework for VLA models, which contains a model-hardware co-aware segmentation strategy and a network-aware deployment adjustment method.
    \item[$\bullet$] We evaluate \textit{RoboECC} through various VLA models and hardware settings, showing its advantages and evaluating its performance in both simulation and real-world scenarios.
\end{itemize}

Experiments show that compared with pure edge-side deployment, \textit{RoboECC} achieves a 3.16$\times$$\sim$3.28$\times$ speedup on the Orin+A100 platform and a 2.10$\times$$\sim$2.23$\times$ speedup on the Thor+A100 platform, with only 2.55\%$\sim$2.62\% overhead.
\section{\textbf{Background}}
\label{tex:background}

\subsection{\textbf{VLA Models}}
\subsubsection{\textbf{Various Model Structure}}
Traditional VLA models~\cite{rt2, openvla} usually consist of three core components: a vision encoder (e.g., Vision Transformer, ViT), a large language model (LLM), and an action detokenizer. 
With technological progress, modern VLA frameworks have increasingly adopted dedicated action models~\cite{cogact} replace simple detokenizers, enabling more precise action generation. 
These models (covering diffusion models, Diffusion Transformer (DiT), LSTM, and MLP) also add complexity to VLA structures.

\subsubsection{\textbf{Inference Cost and Real-Time Requirement}}
The massive parameters and computations of VLA models result in prohibitive inference overhead, failing to meet real-time control demands~\cite{mole-vla}.
Robotic real-time control requires an action generation frequency exceeding 30 Hz, yet unoptimized VLA models only yield 3$\sim$5 Hz, which is drastically below the real-time threshold~\cite{kerv, heisd}.
Existing edge devices are barely capable of real-time VLA inference, making ECC deployment the primary solution for practical VLA applications.

\subsection{\textbf{Edge-Cloud Collaborative Deployment}}

\subsubsection{\textbf{ECC-based Co-Inference}}
Existing ECC deployment frameworks can be broadly categorized into two categories: (1) large-small model co-inference~\cite{spinn, CognitiveECC} and (2) computation offloading~\cite{edgeshard, coedge}.
For large-small model co-inference, common techniques are early-exit mechanism and knowledge distillation.
However, for VLAs, which are characterized by immense size and complexity, it is hard to construct a model supporting early-exit or distilling knowledge.

\subsubsection{\textbf{Computing Offloading for Deployment}}
For computation offloading, the two primary methods are parallelism design~\cite{coedge} and model partitioning~\cite{edgeshard}.
Compared with large-small co-inference methods, computing offloading technologies are more suitable for VLA models.
Therefore, the ECC framework discussed in this paper refers to the computing offloading category.
However, existing ECC frameworks are suboptimal when integrating with VLA models due to the aforementioned reasons in Section~\ref{tex:introduction}.
\section{\textbf{Observation and Analysis}}
\label{tex:analysis}


In this section, we will analyze in detail why existing ECC frameworks do not fit into the VLA model and provide insights on how to design novel ECC frameworks.

\subsection{\textbf{Optimal Segmentation Point under Various Structures}}

\noindent \textit{Motivation \ding{172}: The variable structure of VLAs hinders existing ECC frameworks from determining optimal model segmentation points, undermining ECC deployment performance.}

\begin{figure}[!t]
    \centering
    \includegraphics[width=3.3in]{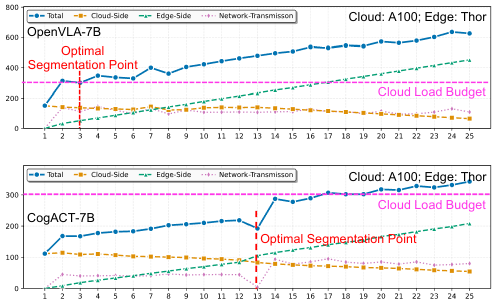}
    \vspace{-2mm}
    \caption{Latency of Model Segmentation under Various Structures}
    \vspace{-4mm}
    \label{fig:2}
\end{figure}

An important part of ECC co-deployment is determining how the original model should be segmented to determine how much load edge-side and cloud-side should bear.
We select two VLA models (OpenVLA~\cite{openvla} and CogACT~\cite{cogact}) with different structures and profile them.
OpenVLA's model structure contains ViT encoders and LLM, without a generative action model.
CogACT's model structure contains ViT encoders, LLM, and a Diffusion-based action model.
We start at the last layer of the model, gradually advance the segmentation points, and record the cloud-side/edge-side latency and network communication latency, shown in Fig.~\ref{fig:2}.

As shown in the Fig.~\ref{fig:2}, OpenVLA does not have an Action Model, so the end structure is an isomorphic LLM Block, and the change in latency is relatively linear.
In this case, given a cloud load budget, the optimal segmentation point can be determined as follows: the block closest to the cloud load budget is near the optimal segmentation point.
However, this approach does not work for other VLA architectures such as CogACT. 
As shown in the Fig.~\ref{fig:2}, although the 16$^{\textnormal{th}}$ block is also under the cloud-side load budget, its latency is significantly higher than the 13$^{\textnormal{th}}$ block. 
In this case, the optimal split point should be the 13$^{\textnormal{th}}$ block.
Therefore, the optimal segmentation point search strategy of VLA needs to be able to adapt to various model structures.

Fortunately, latency increases linearly within the same structure but changes abruptly at structural transitions. 
This suggests two key considerations for latency prediction: models can accurately forecast latency when targeting hardware within a single structure, and they must also incorporate the model structure itself while accounting for structural changes.

\noindent \textit{Insight \ding{172}: While VLA’s diverse structure hinders segmentation point identification, its model structure features certain repeatability, enabling the estimation of a given layer’s hardware execution overhead using hardware performance data.}

\subsection{\textbf{Performance Drift under Network Bandwidth Fluctuation}}

\noindent \textit{Motivation \ding{173}: Network fluctuations can alter transmission delays between the edge and cloud sides, leading to the drift of the optimal segmentation point.}

\begin{figure}[!t]
    \centering
    \includegraphics[width=3.3in]{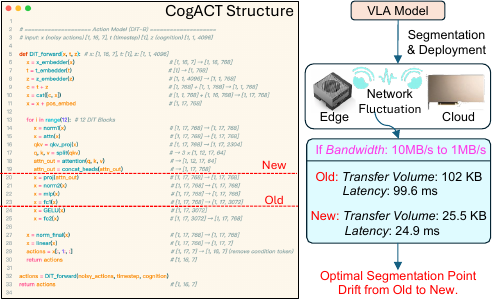}
    \vspace{-2mm}
    \caption{Performance Drift under Network Bandwidth Fluctuation}
    \vspace{-4mm}
    \label{fig:3}
\end{figure}

In ECC deployment, cloud-side and end-side devices connect via a network, while edge-side and cloud-side devices transmit intermediate activations through the network.
Network bandwidth constitutes a non-negligible component of latency, and in certain cases, the proportion of network latency can exceed 40\%.
However, network bandwidth fluctuates in practical scenarios, leading to a drift in the optimal segmentation point.
An example is provided in Fig.~\ref{fig:3}.
Assume that ``Old'' is the optimal segmentation point without considering network fluctuations.
For this segmentation point, the amount of data to be transmitted is $[1, 17, 3072]$ (102 KB).
When the network condition is good (e.g., 10 MB/s), the latency of this segment is only 9.9 ms. 
However, if network fluctuations occur (e.g., a drop from 10 MB/s to 1 MB/s), the latency of this segment increases to 99.6 ms, becoming a bottleneck in edge-cloud deployment.
Therefore, the optimal segmentation point shifts to "New" because the transmission volume at this point decreases to $[1, 17, 768]$ (25.5 KB), with network latency reduced to 24.9 ms and almost no changes in computing latency at both the edge-side and cloud-side.

Existing ECC frameworks designed for LLMs assume that the network transmission delay is static when determining the optimal segmentation point of the model.
This assumption is reasonable for traditional LLMs, as they typically generate relatively long output sequences, which effectively averages out fluctuations in network bandwidth without causing adverse impacts.
Additionally, traditional LLMs do not require multi-step inference and perform only one inference per user request.

Yet, the scenario changes for embodied VLA models.
First, the output of the VLA model is actions, each of which is a very short sequence, thus having no average effect on network fluctuations.
Second, the VLA model relies on multi-step inference to complete the entire task; significant delays in critical steps caused by network fluctuations may lead to severe consequences.
Taking the VLA applied in autonomous driving as an example~\cite{fastdrive-vla}, delays in the generation of critical actions, such as braking, caused by network fluctuations, may lead to traffic accidents.
Thus, the VLA model is highly sensitive to performance drifts caused by network fluctuations, requiring targeted design when deployed in ECC.

\noindent \textit{Insight \ding{173}: The ECC framework for the VLA model requires awareness of network fluctuations and dynamic adjustment of the optimal segmentation point.}

\section{\textbf{\textit{RoboECC} Framework}}
\label{tex:methods}

This section will detail the proposed deployment framework \textit{RoboECC}, which contains two components: (1) model-hardware co-aware segmentation strategy and (2) network-aware deployment adjustment approach.

\subsection{\textbf{Model-Hardware Co-Aware Segmentation Strategy}}
\label{tex:methods-1}

\subsubsection{\textbf{Structure Modeling of Various VLA Models}}
We first built a structure modeling for VLA models.
We divide the VLA models into three sets, namely encoder $\mathbb{S}_{\textnormal{enc}}$, backbone $\mathbb{S}_{\textnormal{bac}}$, and decoder $\mathbb{S}_{\textnormal{dec}}$.
For a given VLA model, its structure can be expressed as $[\mathbb{S}_{\textnormal{enc}}, \mathbb{S}_{\textnormal{bac}}, \mathbb{S}_{\textnormal{dec}}]$.
Then, we specify that:
$\mathbb{S}_{\textnormal{enc}} \in \{ \mathcal{M}_{\textnormal{ViT}} \}$; $\mathbb{S}_{\textnormal{bac}} \in \{ \mathcal{M}_\textnormal{LLM} \}$ and $\mathbb{S}_{\textnormal{dec}} \in \{ \mathcal{M}_\textnormal{De-tokenizer}, \mathcal{M}_\textnormal{MLP}, \mathcal{M}_\textnormal{LSTM}, \mathcal{M}_\textnormal{Diff}, \mathcal{M}_\textnormal{DiT} \}$.
These cover all current VLA model structures.

Further, we test the fine-grained layer category $L_{i}$ in each type of structure, the hidden size of each layer $H_{i}^{L}$, and the number of parameters $W_{i}$ to establish a mapping.
We actually measure the total amount of computation and data movements required for each of these structures, represented as $\mathcal{C}^{type}_{compute}$ (FLOPs) and $\mathcal{C}^{type}_{datamove}$ (KB).
When the model structure and details are clear, searching based on this mapping can obtain the cost details for each part, shown in Eq.~\ref{eq:1}.
\begin{equation}
\textnormal{Search in} \ \mathcal{M}_{type}(L_i,H_{i}^{L}, W_{i}) \rightarrow (\mathcal{C}_{compute}^{type}, \mathcal{C}_{datamove}^{type} ).
\label{eq:1}
\end{equation}

\begin{algorithm}[!b]
\footnotesize
\caption{Automatic Searching Algorithm to Find the Optimal Model Segmentation Point}
\begin{algorithmic}[1]
\Require Edge-Side GPU Latency $T_{GPU}^{Edge}(L_{i})$, Cloud-Side $T_{GPU}^{Cloud}(L_{i})$, Cloud Offload Budget $B^{Cloud}$, Model Segmentation Point $S$, Number of Layers $n$.
\Ensure Cloud Offload $load^{Cloud}(\mathcal{C}^{L{_i}}_{compute}; \mathcal{C}^{L_{i}}_{memory})$.
    \For{Each $S$ } $T_{total}^{S} = \sum_{0}^{S} T_{GPU}^{Edge}(L_{i}) + \sum_{S}^{n} T_{GPU}^{Cloud}(L_{i})$ 
        \If{$ \sum_{S}^{n} load^{cloud}(\mathcal{C}^{L{_i}}_{compute}; \mathcal{C}^{L_{i}}_{memory}) < B^{cloud}$}
        \State $ S = S+1 $ 
        \Else $\ \textnormal{\textbf{break}}$
        \EndIf
    \If{$T_{total}^{S} < T_{total}^{\textnormal{optim}}$}
    \State Optim Point: $ P_{\textnormal{optim}} = S $ and $T_{total}^{\textnormal{optim}} = T_{total}^{S}$
    \Else $ \  \textnormal{Maintain} \ P_{\textnormal{optim}} $
    \EndIf
    \EndFor
\end{algorithmic}
\label{alg:1}
\end{algorithm}

\subsubsection{\textbf{Hardware Modeling and Latency Computation}}
However, it is not enough to obtain latency only by the structure modeling, because the latency is also related to hardware performance.
Therefore, we propose a computational modeling for both edge-side/cloud-side GPU hardware performance.
Considering GPUs' pipeline design, the theoretical latency can be calculated by Eq.~\ref{eq:2}.
\begin{equation}
\begin{aligned}
T_{GPU} & = \sum_{i=0}^{L_{i}}max(T_{compute}^{L{i}}, T_{memory}^{L{i}})\\
& = \sum_{i=0}^{L_{i}}max(\frac{\mathcal{C}_{compute}^{L_{i}}}{P_{i}\times Parallel_{i}}, \frac{\mathcal{C}_{datamove}^{L_{i}}}{Bandwidth_{i}}),
\end{aligned}
\label{eq:2}
\end{equation}
where $L_{i}$ is the layer type, $ P_{i} $ is the computing power of the specified GPU, and $ Bandwidth_{i} $ indicates the hardware bandwidth between the GPU and the global memory.
To date, structural and hardware modeling are finished, with both utilized for calculating latency.

\subsubsection{\textbf{Optimal Model Segmentation}}
Based on the aforementioned modeling, we can automatically determine the optimal segmentation point of the model.
We perform a depth-first search in a given search space.
Note that the search space is determined by the load budget given by the cloud device.
The overall search process is shown in Alg.~\ref{alg:1}.
Specifically, we start from the last layer and identify the optimal segmentation point within the allowable cloud-side load range to minimize overall computational delay. 
We do not integrate a more advanced search strategy into the algorithm because all computational data is derived from our modeling rather than actual measurements or fitting data. This results in extremely low computational load for the search algorithm, with negligible overhead that requires no additional optimization.

\begin{figure}[!t]
    \centering
    \includegraphics[width=3.3in]{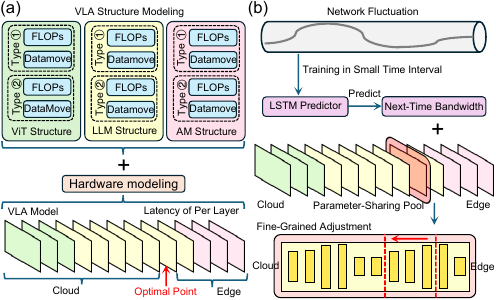}
    \vspace{-2mm}
    \caption{Components in \textit{RoboECC}: (a) Model-Hardware Co-Aware Segmentation Strategy (2) Network-Aware Deployment Adjustment Approach}
    \vspace{-4mm}
    \label{fig:4}
\end{figure}

\subsection{\textbf{Network-Aware Deployment Adjustment Approach}}
\label{tex:methods-2}

\subsubsection{\textbf{Network Fluctuation Predictor}}
To dynamically perceive changes in network bandwidth, we design network fluctuation predictors simultaneously on the cloud and edge sides. 
Long-term historical data of network bandwidth between the edge and cloud sides is collected to train a lightweight LSTM network.
It is important to note that, as our requirement is to predict network bandwidth in real time, the LSTM model needs to be trained with a very fine granularity.
It is necessary to ensure that the time interval of input data $t_{input}$ in the LSTM is shorter than the minimum delay of each component after VLA segmentation, as shown in Eq.~\eqref{eq:3}.
\begin{equation}
\{t_{\textnormal{input}} \ | \ LSTM_{train}\}< min(t_{\textnormal{cloud}}, t_{\textnormal{edge}})
\label{eq:3}
\end{equation}

\subsubsection{\textbf{Parameter-Sharing Pool}}
To enable network-aware tuning, dynamic adjustments to parameters and model structure are essential. 
However, transmitting these parameters over the network would introduce substantial latency. 
To address this challenge, we propose an innovative parameter-sharing pool. 
Specifically, we store all layers of the block containing the optimal segmentation point on both the cloud and edge sides. 
Since only a single block is involved, the weight overhead for storing this component remains relatively low. 
Taking the LLAMA Block as an example, the parameters of a layer occupy merely approximately 386 MB. 
For the cloud, the primary load lies in computational/runtime overhead. Therefore, 386 MB of memory is negligible in this context.
Meanwhile, modern edge-side devices are typically equipped with large memory capacities (e.g., Orin with 32 GB / 64 GB and Thor with 64 GB / 128 GB), which are rarely fully utilized by VLA models (for instance, OpenVLA consumes only around 16 GB of memory). 
Thus, placing these sharing parameters on the edge and cloud sides does not incur significant overhead.

\subsubsection{\textbf{Fine-Grained Segmentation Adjustment}}
We calculate the variation of network bandwidth $\Delta NB = NB_{t+1}^{\textnormal{pre}} - NB_{t}^{\textnormal{real}}$ based on the next-moment network bandwidth predicted by LSTM.
We set two thresholds $\mathcal{T}_{high}$ and $\mathcal{T}_{low}$ to determine whether dynamic adjustments are required.
When $\Delta NB$ is higher than $\mathcal{T}_{high}$, the network bandwidth increases; within the parameter-sharing pool, we search for the layer with the maximum amount of data to be transmitted and adjust the segmentation point to this layer to maximize the utilization of network bandwidth. 
When $\Delta NB$ is lower than $\mathcal{T}_{low}$, the network bandwidth decreases; in this case, we search for the layer with the minimum amount of data to be transmitted within the parameter-sharing pool and adjust the segmentation point to this layer to minimize the data transmission volume.
These fine-grained adjustments do not require considering changes in computational load on the edge or cloud side, as their impacts on both sides are negligible.
\section{\textbf{Experiments}}
\label{tex:experiment}
\vspace{-1.5mm}
\subsection{\textbf{Setup}}
\label{tex:expriment-1}

\begin{figure*}[!t]
    \centering
    \includegraphics[width=6.7in]{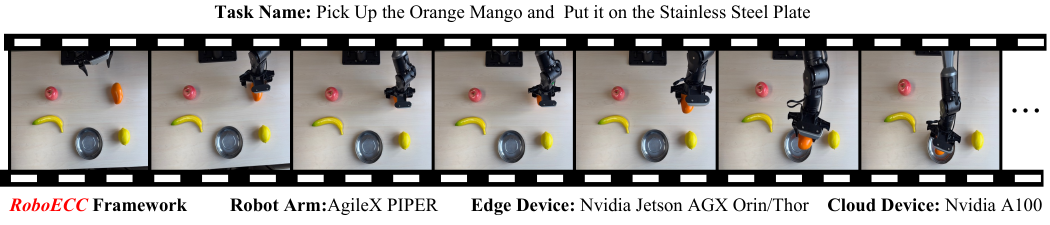}
    \vspace{-2mm}
    \caption{An Example of \textit{RoboECC} Deployment in Real-World Scenarios}
    \vspace{-4mm}
    \label{fig:5}
\end{figure*}

\begin{figure}[!t]
    \centering
    \includegraphics[width=3in]{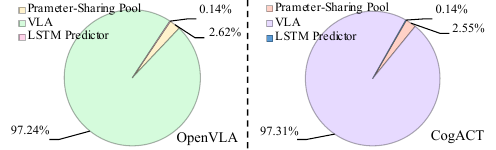}
    \vspace{-1mm}
    \caption{Overhead of the proposed \textit{RoboECC} Framework}
    \vspace{-4mm}
    \label{fig:6}
\end{figure}

\begin{figure}[!t]
    \centering
    \includegraphics[width=3in]{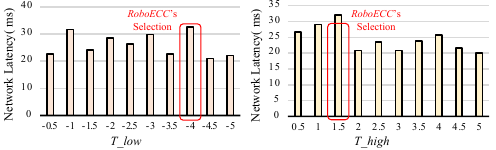}
    \vspace{-2mm}
    \caption{Hyper-parameters $\mathcal{T}_{low}$ and $\mathcal{T}_{high}$ in \textit{RoboECC}}
    \vspace{-5mm}
    \label{fig:7}
\end{figure}

\subsubsection{\textbf{Hardware Platform}}
We select the Nvidia A100 GPU as the cloud-side device and the Nvidia AGX Jetson Orin/Thor as edge-side devices as platforms, the details of which are shown in Tab.~\ref{tab:hardware}.
The power consumption modes of Orin and Thor are uniformly set to MAX\_N mode, as VLA models require high real-time performance, and the maximum power consumption mode can reduce latency.

\begin{table}[!b]
    \centering
    \vspace{-5mm}
    \scriptsize
    \caption{Details of Cloud/Edge-Side Hardware Platform.}
    \vspace{-1mm}
    \begin{tabular}{c|c|c|c}
    \toprule
    \toprule
    ~ & \textbf{Cloud-Side} & \multicolumn{2}{c}{\textbf{Edge-Side}} \\
    \midrule
    \textbf{Platform} & A100 GPU & Jetson Orin & Jetson Thor \\
    \midrule
    \textbf{Frequency} & 1410 MHz & 1300 MHz & 1386 MHz \\
    \midrule
    \textbf{Computing Power (4-bit)} & 2496 TFLOPs & 275 TFLOPs & 517.5 TFLOPs \\
    \midrule
    \textbf{Tensor Core Num.} & 432 & 64 & 96 \\
    \midrule
    \textbf{Memory Bandwidth} & 2039 GB/s & 204.8 GB/s & 273 GB/s \\
    \bottomrule
    \bottomrule
    \end{tabular}
    \label{tab:hardware}
\end{table}

\subsubsection{\textbf{Models}}
We select OpenVLA~\cite{openvla} as the evaluation model. 
It generates action sequences token-by-token directly through its language model backbone, without employing independent action modules. 
OpenVLA provides a controlled baseline to isolate and evaluate temporal-dynamic sensitivity during sequential inference.

\subsubsection{\textbf{Baselines}}
Since existing ECC frameworks are not suitable for VLA model deployment, we use device-side-only deployment and cloud-side-only deployment as the baselines.
Moreover, we establish a fixed segmentation scheme (with the load approximately equally distributed between the edge and cloud sides) and compare it with \textit{RoboECC}.
Comparison with these baselines well demonstrates the advantages of \textit{RoboECC}.

\subsubsection{\textbf{Evaluation Metrics}}
We adopt edge-side latency, cloud-side latency, total latency, cloud-side load, and edge-side load as the core evaluation metrics.
Notably, \textit{RoboECC} does not affect the task success rate, which is therefore not reported.

\subsection{\textbf{Main Results}}
\label{tex:expriment-2}
\subsubsection{\textbf{Results in Simulation Benchmark}}
Tests are conducted in the LIBERO~\cite{libero} and SimplerEnv~\cite{simplerenv} simulation benchmark.
We conduct 10 tests on each simulation task and report the average inference latency of each step.
The results are shown in Tab.~\ref{tab:simulation-openvla} (OpenVLA) and Tab.~\ref{tab:simulation-cogact} (CogACT).
Note that total latency includes network latency.
Compared with pure edge-side deployment, \textit{RoboECC} achieves a 3.16$\times$$\sim$3.28$\times$ speedup on the Orin+A100 platform and a 2.10$\times$$\sim$2.23$\times$ speedup on the Thor+A100 platform.
Compared with pure cloud-side deployment, \textit{RoboECC} reduces cloud-side load effectively without introducing excessive inference latency.
Moreover, compared with fixed segmentation schemes, \textit{RoboECC} exhibits significantly better edge-cloud trade-offs.
These results show the superiority of \textit{RoboECC}.
\begin{table}[!t]
\centering
\scriptsize
\setlength{\tabcolsep}{1mm}
\caption{\textit{RoboECC}'s Performance under OpenVLA LIBERO}
\vspace{-2mm}
\label{tab:simulation-openvla}
\begin{tabular}{c|c|c|c|c|c|c|c}
\toprule
\toprule
\multicolumn{8}{c}{\textbf{\textcolor{blue}{OpenVLA + LIBERO} in \textit{RoboECC}}} \\
\midrule
\multirow{2}{*}{\textbf{HW}} & \multirow{2}{*}{\textbf{Methods}} & \multicolumn{2}{c|}{\textbf{Cloud-Side}} & \multicolumn{2}{c|}{\textbf{Edge-Side}} & \multicolumn{2}{c}{\textbf{Total}} \\
\cmidrule{3-8}
~ & ~ & \textbf{Lat.} & \textbf{Load} & \textbf{Lat.} & \textbf{Load} & \textbf{Lat.} & \textbf{Load}  \\
\midrule

\multirow{4}{*}{Orin} & \cellcolor{blue!20}{Edge-Only} & \cellcolor{blue!20}{--} & \cellcolor{blue!20}{--} & \cellcolor{blue!20}{1119.4ms} & \cellcolor{blue!20}{14.1GB} & \cellcolor{blue!20}{1119.4ms} & \cellcolor{blue!20}{14.1GB} \\

~ & \cellcolor{red!20}{Cloud-Only} & \cellcolor{red!20}{151.2ms} & \cellcolor{red!20}{14.1GB} & \cellcolor{red!20}{--} & \cellcolor{red!20}{--} & \cellcolor{red!20}{151.2ms} & \cellcolor{red!20}{14.1GB} \\

~ & \cellcolor{orange!20}{Fixed Seg} & \cellcolor{orange!20}{87.9ms} & \cellcolor{orange!20}{6.3GB} & \cellcolor{orange!20}{717.8ms} & \cellcolor{orange!20}{7.9GB} & \cellcolor{orange!20}{923.3ms} & \cellcolor{orange!20}{14.2GB}\\

~ & \cellcolor{green!20}{\textit{RoboECC}} & \cellcolor{green!20}{136.7ms} & \cellcolor{green!20}{12.1GB} & \cellcolor{green!20}{94.5ms} & \cellcolor{green!20}{2.0GB} & \cellcolor{green!20}{354.4ms} & \cellcolor{green!20}{14.1GB}\\ 
\midrule
\multirow{4}{*}{Thor} & \cellcolor{blue!20}{Edge-Only} & \cellcolor{blue!20}{--} & \cellcolor{blue!20}{--} & \cellcolor{blue!20}{628.9ms} & \cellcolor{blue!20}{14.1GB} & \cellcolor{blue!20}{628.9ms} & \cellcolor{blue!20}{14.1GB} \\

~ & \cellcolor{red!20}{Cloud-Only} & \cellcolor{red!20}{151.2ms} & \cellcolor{red!20}{14.1GB} & \cellcolor{red!20}{--} & \cellcolor{red!20}{--} & \cellcolor{red!20}{151.2ms} & \cellcolor{red!20}{14.1GB} \\

~ & \cellcolor{orange!20}{Fixed Seg} & \cellcolor{orange!20}{89.5ms} & \cellcolor{orange!20}{6.3GB} & \cellcolor{orange!20}{378.4ms} & \cellcolor{orange!20}{7.9GB} & \cellcolor{orange!20}{587.2ms} & \cellcolor{orange!20}{14.2GB}\\

~ & \cellcolor{green!20}{\textit{RoboECC}} & \cellcolor{green!20}{137.1ms} & \cellcolor{green!20}{12.1GB} & \cellcolor{green!20}{51.3ms} & \cellcolor{green!20}{2.0GB} & \cellcolor{green!20}{300.1ms} & \cellcolor{green!20}{14.1GB}\\ 
\bottomrule
\bottomrule
\end{tabular}
\vspace{-4mm}
\end{table}
\begin{table}[!b]
\centering
\vspace{-5mm}
\scriptsize
\setlength{\tabcolsep}{1mm}
\caption{\textit{RoboECC}'s Performance under CogACT+SimplerEnv}
\vspace{-1mm}
\label{tab:simulation-cogact}
\begin{tabular}{c|c|c|c|c|c|c|c}
\toprule
\toprule
\multicolumn{8}{c}{\textbf{\textcolor{blue}{CogACT + SimplerEnv} in \textit{RoboECC}}} \\
\midrule
\multirow{2}{*}{\textbf{HW}} & \multirow{2}{*}{\textbf{Methods}} & \multicolumn{2}{c|}{\textbf{Cloud-Side}} & \multicolumn{2}{c|}{\textbf{Edge-Side}} & \multicolumn{2}{c}{\textbf{Total}} \\
\cmidrule{3-8}
~ & ~ & \textbf{Lat.} & \textbf{Load} & \textbf{Lat.} & \textbf{Load} & \textbf{Lat.} & \textbf{Load}  \\
\midrule

\multirow{4}{*}{Orin} & \cellcolor{blue!20}{Edge-Only} & \cellcolor{blue!20}{--} & \cellcolor{blue!20}{--} & \cellcolor{blue!20}{775.3ms} & \cellcolor{blue!20}{14.5GB} & \cellcolor{blue!20}{775.3ms} & \cellcolor{blue!20}{14.5GB} \\

~ & \cellcolor{red!20}{Cloud-Only} & \cellcolor{red!20}{111.4ms} & \cellcolor{red!20}{14.5GB} & \cellcolor{red!20}{--} & \cellcolor{red!20}{--} & \cellcolor{red!20}{111.4ms} & \cellcolor{red!20}{14.5GB} \\

~ & \cellcolor{orange!20}{Fixed Seg} & \cellcolor{orange!20}{46.9ms} & \cellcolor{orange!20}{7.9GB} & \cellcolor{orange!20}{437.2ms} & \cellcolor{orange!20}{6.6GB} & \cellcolor{orange!20}{572.5ms} & \cellcolor{orange!20}{14.5GB}\\

~ & \cellcolor{green!20}{\textit{RoboECC}} & \cellcolor{green!20}{81.9ms} & \cellcolor{green!20}{12.0GB} & \cellcolor{green!20}{143.2ms} & \cellcolor{green!20}{2.5GB} & \cellcolor{green!20}{236.1ms} & \cellcolor{green!20}{14.5GB}\\ 
\midrule
\multirow{4}{*}{Thor} & \cellcolor{blue!20}{Edge-Only} & \cellcolor{blue!20}{--} & \cellcolor{blue!20}{--} & \cellcolor{blue!20}{429.6ms}  & \cellcolor{blue!20}{14.5GB} & \cellcolor{blue!20}{429.6ms} & \cellcolor{blue!20}{14.5GB} \\

~ & \cellcolor{red!20}{Cloud-Only} & \cellcolor{red!20}{111.4ms} & \cellcolor{red!20}{14.5GB} & \cellcolor{red!20}{--} & \cellcolor{red!20}{--} & \cellcolor{red!20}{111.4ms} & \cellcolor{red!20}{14.5GB} \\

~ & \cellcolor{orange!20}{Fixed Seg} & \cellcolor{orange!20}{47.2ms} & \cellcolor{orange!20}{7.9GB} & \cellcolor{orange!20}{240.4ms} & \cellcolor{orange!20}{6.6GB} & \cellcolor{orange!20}{375.4ms} & \cellcolor{orange!20}{14.5GB}\\

~ & \cellcolor{green!20}{\textit{RoboECC}} & \cellcolor{green!20}{82.7ms} & \cellcolor{green!20}{12.0GB} & \cellcolor{green!20}{105.7ms} & \cellcolor{green!20}{2.5GB} & \cellcolor{green!20}{192.7ms} & \cellcolor{green!20}{14.5GB}\\ 
\bottomrule
\bottomrule
\end{tabular}
\end{table}

\vspace{-2.5mm}
\subsubsection{\textbf{Results in Real-World Environment}}
We conduct tests on real robotic systems using OpenVLA.
We use the AgileX PIPER robotic arm and Orin+A100/Thor+A100 as the platform.
We collect 1,000 data samples for model fine-tuning, following which the CogACT and OpenVLA are deployed using the \textit{RoboECC} framework.
A practical example is presented in Fig.~\ref{fig:5}.
When this task is executed, the inference latency of edge-only deployment is approximately 1274.4 ms (Orin) and 667.2 ms (Thor).
When integrating \textit{RoboECC}, it only needs 361.8 ms (Orin+A100) and 321.6 ms (Thor+A100), making the movement of the robotic arm smoother and reducing mechanical jitter.
The proposed \textit{RoboECC} adapts to real-world scenarios, demonstrating robustness.

\subsubsection{\textbf{Ablation Studies}}
Ablation experiments are conducted on the Orin+A100 platform, with the results presented in Tab.~\ref{tab:ablation}.
The integration of co-aware segmentation leads to a significant reduction in overall latency, attributed to the performance gains provided by the cloud side.
Incorporating network-aware adjustment results in negligible changes in cloud-side and edge-side latency but a notable decrease in network latency, thereby further lowering the total latency.

\begin{table}[!b]
    \centering
    \vspace{-5mm}
    \scriptsize
    \caption{Ablation Studies of \textit{RoboECC}.}
    \vspace{-2mm}
    \begin{tabular}{c|c|c|c}
    \toprule
    \toprule
    \textbf{Methods} & \textbf{Cloud-Side} & \textbf{Edge-Side} & \textbf{Total} \\
    \midrule
    \textbf{Edge-Only} & -- & 1119.4ms & 1119.4ms\\
    \textbf{+ Co-Aware Segmentation} & 135.9ms & 99.8ms & 392.7ms \\
    \textbf{+ Network-Aware Adjustment} & 136.7ms & 94.5ms & 354.4ms \\
    \bottomrule
    \bottomrule
    \end{tabular}
    \label{tab:ablation}
\end{table}

\subsection{\textbf{Discussion}}
\label{tex:expriment-3}

\subsubsection{\textbf{Overhead}}

In \textit{RoboECC}, a lightweight LSTM is employed as the network bandwidth predictor, which measures merely 20.1 MB in size. 
This overhead is negligible compared to that of the VLA model. 
As illustrated in Fig.~\ref{fig:6}, the parameter-sharing pool accounts for only 2.55\%$\sim$2.62\% of the total overhead, while the LSTM-related overhead is negligible.
We test the time overhead of split point adjustment under different network bandwidths, and the results show that its average value is only 10.7 ms, while the average latency reduction achieved is 32.6 ms, indicating that the overhead of this part is far less than the gain.

\subsubsection{\textbf{Hyper-Parameters}}
The hyper-parameters $\mathcal{T}_{low}$ and $\mathcal{T}_{high}$ are determined using the following method:
We take the maximum $\Delta NB$ in historical network bandwidth changes as $\mathcal{T}_{high}$, then conduct experiments to determine $\mathcal{T}_{low}$. 
After $\mathcal{T}_{low}$ is determined, we further perform experiments to update $\mathcal{T}_{high}$.
The results of this process are illustrated in Fig.~\ref{fig:7}.

\subsubsection{\textbf{Scope}}
In \textit{RoboECC}, we only provide a hardware model for GPUs, and have not yet considered accelerators with other architectures, which are regarded as future work.
\vspace{-1mm}
\section{\textbf{Conclusion}}
\label{tex:conclusion}

We propose a multi-factor-aware ECC deployment framework for VLA models, named \textit{RoboECC}.
In \textit{RoboECC}, we propose a model-hardware co-aware segmentation strategy and a network-aware deployment adjustment approach for optimal performance.
Experiments show that \textit{RoboECC} achieves a speedup of up to 3.28x with only 2.55\%$\sim$2.62\% overhead.

\bibliographystyle{IEEEtran}
\bibliography{reference/ref.bib}

\end{document}